\documentclass[iop]{emulateapj-rtx4}

\shorttitle{Long-term Chandra variability survey of  M31; $\sim$200 new XBs}
\shortauthors{Barnard et al.}

\begin{document}


\title{Around 200 new X-ray binary IDs from 13 years of Chandra observations of the M31 center}


\author{R. Barnard, M. Garcia, and F. Primini}
\affil{Harvard-Smithsonian Center for Astrophysics, Cambridge, MA 02138}
\and
\author{Z. Li}
\affil{School of Astronomy and Space Science, Nanjing University, Nanjing 210093, China}
\and
\author{F. K. Baganoff}
\affil{Center for Space Research, M.I.T., Cambridge, MA 02139} 
\and
\author{S. S. Murray}
\affil{Johns Hopkins University, Baltimore, Maryland}


\begin{abstract}
We have created 0.3--10 keV, 13 year, unabsorbed  luminosity lightcurves for 528 X-ray sources in the central 20$'$ of M31. We have 174 Chandra observations spaced at $\sim$1 month intervals thanks to our transient monitoring program,  deeper observations of the M31 nucleus, and some public data from other surveys. We created 0.5--4.5 keV structure functions (SFs) for each source, for comparison with the ensemble structure function of AGN. We find 220 X-ray sources with luminosities $\ga$10$^{35}$ erg s$^{-1}$  that have SFs with significantly more variability than the ensemble AGN SF, and are  likely X-ray binaries (XBs). A further 30 X-ray sources were identified as XBs using other methods. We therefore have 250 probable XBs in total, including $\sim$200 new identifications. This result represents great progress over the $\sim$50 XBs and $\sim$40 XB candidates previously identified out of the $\sim$2000 X-ray sources within the D$_{25}$ region of M31; it also demonstrates the power of SF analysis for identifying XBs in external galaxies. 
We also identify a new transient black hole candidate, associated with the M31 globular cluster B128.
\end{abstract}





\section{Introduction}

The X-ray populations of external galaxies have the potential to provide excellent diagnostics on the evolutionary states of their host galaxies. However, separating the true galaxy population from active galaxies in the field is notoriously difficult. This is because X-ray binaries (XBs) are expected to dominate the X-ray populations of galaxies, and their emission spectra can be very similar to the spectra of active galactic nuclei (AGN). In fact, it has been rather difficult to classify X-ray sources in external galaxies at all;  \citet{stiele11} examined 1897 X-ray sources within the D$_{25}$ region for M31, but only identified 46 XBs and 43 XB candidates, while $\sim$65\% of the X-ray sources had no classification. However, we have recently invented a method for discriminating between XBs and AGN by formalizing their differences in variability \citep{barnard2012c}.

 It has long been known that AGN may vary by a factor of  2--3 on time-scales of months to years, with the amplitude of variation inversely proportional to the luminosity \citep[see e.g.][ and references within]{marshall81,nandra97}.  However, exceptional AGN can  flare   up by an order of magnitude \citep[e.g.][]{tan78}.

 Recently \citet{vagnetti11} studied the ensemble variability over time-scales of hours to years of AGN in the serendipitous source catalogs from XMM-Newton \citep{watson09} and Swift \citep{puccetti11}; their sample covered red-shifts $\sim$0.2--4.5, and  0.5--4.5 keV luminosities $\sim$10$^{43}$--10$^{46}$ erg s$^{-1}$.  They included 412 AGN from the XMM-Newton catalog, and 27 AGN from the Swift catalog; all of these AGN were sampled at least twice. They used a structure function (SF) to estimate the mean intensity deviation for data separated by time $\tau$:
\begin{equation}
SF\left( \tau \right) \equiv \sqrt{\frac{\pi}{2}\left<|\log f_{\rm X} \left( t+\tau \right) - \log f_{\rm X} \left(t\right)| \right>^2 - \sigma_{\rm n}^2},
\end{equation}
where $\sigma_{\rm n}$ is the photon noise and $f_{\rm X}$ is the X-ray flux. They grouped  the SF into logarithmic bins with width 0.5; each bin in the range log($\tau$) = 0.0--3.0 contained more than 100 measurements.

 \citet{vagnetti11} found good agreement between the XMM-Newton and Swift samples of AGN, after the noise components were subtracted; SF($\tau$) $\propto$ $\tau^{0.10\pm0.01}$ for the XMM-Newton sample, and SF($\tau$) $\propto$ $\tau^{0.07\pm0.04}$ for the Swift sample. They also investigated the well-observed anti-correlation between intensity variability ($I_{\rm var}$)  and luminosity, expressed  in the form $I_{\rm var} \propto L_{\rm X}^{-k}$; $k$ $\sim$0.3 in the literature for time-scales of days to tens of days. They measured $k$ for AGN grouped logarithmically over log $L_{\rm X}$ = 43.5--45.5 for two values of $\tau$: 1 day and 100 days. They find $k$ = 0.42$\pm$0.03 for $\tau$ = 1 day, and $k$ = 0.21$\pm$0.07 for $\tau$ = 100 days; the latter result is consistent with the work of \citet{markowitz04}, who found $k$ $\sim$0.13 for variations over year-long time-scales. 


Recently, \citet{stiele08} examined the variability of X-ray sources in the M31 center over 9 deep XMM-Newton observations spanning 2000 June -- 2004 July; they obtained fluxes or upper limit for each X-ray source in every observation, and searched for differences between maximum and minimum flux $>$3$\sigma$. They found 149 sources out of 300 to exhibit $>3\sigma$ flux  variability; of those, 44 exhibited variability by a factor $>$5, including 28 XBs or XB candidates.  \citet{stiele08} reclassified any ``hard'' X-ray source which varied by a factor $>$10 as a candidate XB; however, according to the SF obtained by \citet{vagnetti11}, typical AGN only vary by  $\la$70\% over 4 years, meaning that further XBs could have been identified in hind-sight from that variability study.

We have been studying the long-term variability of 528 X-ray sources over 174 Chandra observations of the inner regions of M31, taken over $\sim$13 years. 
We conducted a pilot structure function survey, on 37 X-ray sources associated with globular clusters that are probable low mass XBs \citep{barnard2012c}. We found that the lower  luminosity XBs tended to exhibit considerably more variability than the ensemble AGN SF; the higher luminosity XBs exhibited SFs that indicated comparable or lesser variability than the ensemble AGN SF. Such dependence on variation with luminosity is well known in Galactic systems, where the Z-sources (high luminosity XBs) vary only by a factor of a few, while the atoll sources (low luminosity XBs) vary by $\sim$1--2 orders of magnitude \citep{muno02}.

In this work we will present an overview of the long-term variability of the 528 X-ray sources in our chosen region, out to 20$'$ ($\sim$4.5 kpc) from the center of M31.  We have  created 0.5--4.5 keV SFs for each target, for comparison with the ensemble AGN SF created by \citet{vagnetti11}. With these results, we will place limits on the numbers of XBs and AGN in this region. We will also compare the number of possible  AGN with the expected number from the 0.5--10 keV AGN flux distribution created by \citet{georgakakis08}.

We have previously published  detailed studies of 13  X-ray transients with Chandra and HST coverage \citep{barnard2012b, barnard13c}, and of those X-ray sources associated with M31 globular clusters \citep{barnard2012c}. We also recently published our 26 new black hole candidates that we identified via structure function analysis \citep{barnard13}. Further work on the $\sim$50 transients in our study will be presented separately (R. Barnard et al., in prep). In the following sections, we provide details of the observations and data reduction, followed by results and our discussion.


\section{Observations and data reduction}

The central region of M31 has been observed with Chandra on a $\sim$ monthly basis for the last $\sim$13 years in order to monitor transients; we exclude periods when M31 cannot be viewed due to orbital constraints (approximately March--May each year). We have analyzed 112 ACIS observations and 62 HRC observations, in order to discern the variability of X-ray sources in this region.

Initial source detection was performed on  a merged ACIS events file created from observations performed in 1999 October--2010 March; the detection procedure followed that outlined by \citet{wang04}. Each observation was registered to a single coordinate system, with the systematic uncertainty in registration included in the uncertainty in the centroid position in the merged image. The total exposure was 305 ks but the combined effective 0.5--8 keV exposure  varied considerably over the field of view, from $\sim$2$\times 10^5$ s in the center to $\sim$1.1$\times 10^5$ s within 6$'$ of the center, and down to $\sim10^4$ ks at larger off-axis angles. The highest sensitivity was found $\sim$2$'$ from the nucleus ($\sim 7\times 10^{-5}$ count s$^{-1}$, or $\sim 5\times 10^{34}$ erg s$^{-1}$); diffuse emission is highly significant within 2$'$, while a combination of degrading PSFs and lower exposures decreased the sensitivity by a factor $\ga$20 at high off-axis angles.

Sources were considered significant if they satisfied one or both of the following criteria. A source was considered significant if the best fit line of constant intensity was $\ge$3$\sigma$ above zero; some transient X-ray sources would be rejected by this criterion, so we also accepted sources that demonstrated significant variability.

We compared our initial Chandra source list with the inventory from an XMM-Newton survey of M31 performed by \citet{stiele11}. We found that the XMM-Newton catalog contained  sources that were not in our initial sample. We obtained lightcurves for each of those sources, and any that were significantly detected in our observations were added to the source list.

We determined the position of each source from a merged 0.3--7 keV ACIS image, using the {\sc iraf} tool {\sc imcentroid}. This merged image is registered to the B band image of M31 Field 5 in the Local Group Survey of \citet{massey06} using 27 X-ray-bright globular clusters, as described in \citet{barnard2012b}. For transients that were more recent than the merged Chandra image, we registered the Chandra observation with the highest X-ray flux for that transient to the merged Chandra image. For most sources, the position uncertainties combine the systematic uncertainty in registering the merged Chandra image to the Field 5 B band image, and the statistical uncertainty in the position of the X-ray centroid; new transients have an additional uncertainty in registering the peak observation to the merged image.

We obtained 0.3--7.0 keV  spectra from circular source and background regions for each source. The background region was the same size as the source region, and at a similar off-axis angle. The extraction radius varied between sources, because  larger off-axis angles resulted in larger point spread functions.

\subsection{Converting from intensity to luminosity}
We used XSPEC to convert from 0.3--7.0 keV intensity to 0.3--10 keV luminosity for each observation of every source. Rather than assuming a single conversion factor for a particular emission model, we calculated the conversion factor at the location of each X-ray source in every observation; this was necessary because any given source could be observed in several different parts of the detector, at various off-axis angles, as the roll angle changed between observations.

 We expect most of our X-ray sources to be XBs or AGN; furthermore, most of the XBs are likely to be in the ``hard state'' common to all XBs \citep{vdk94} at luminosities $\la$10\% Eddington \citep{gladstone07, tang11}. The emission spectra of such sources may be estimated by a power law with photon index ($\Gamma$) of  1.7, hence we initially assumed an absorbed power law model for most X-ray sources, with $\Gamma$ = 1.7 and line-of-sight absorption ($N_{\rm H}$) equivalent to 7$\times 10^{20}$ atom cm$^{-2}$, the Galactic column density in the direction of M31 \citep{stark92}. However, stars and supersoft sources were modeled by 0.05 keV black body spectra with $N_{\rm H}$ = 7$\times 10^{20}$ atom cm$^{-2}$.

\subsubsection{ACIS}
  For ACIS observations, we obtained the response matrices and ancillary response files corresponding to each source spectrum. Net source spectra with $>$200 photons were freely fitted, starting with the assumed spectra, but with $N_{\rm H}$ and $\Gamma$ free to vary; if the best fit $N_{\rm H}$ was $<$ 7$\times 10^{20}$ atom cm$^{-2}$, then we fixed $N_{\rm H}$ to 7$\times 10^{20}$ atom cm$^{-2}$. If an X-ray source exhibited spectra with $>$200 net counts in only one observation, then we obtained a new lightcurve assuming the best fit model for that observation. If a source had $>$200 source counts in more than one spectrum, then we plotted $N_{\rm H}$ and $\Gamma$ (or $kT$) vs. time, and found the best fit line of constant parameter ($N_{\rm H}$, $\Gamma$ or $kT$), then converted from intensity to flux using these best fit values for observations with $<$200 net counts. 

 For transient X-ray sources, we tried two emission models: a power law representing the hard state, and a disk blackbody representing the black hole high state that is often observed in black hole transients during outburst \citep{remillard06}. If the disk blackbody model was more successful, then we only applied this model to the outburst.

We took this approach because it is unfeasible to fit the individual spectra simultaneously, or create merged spectra for each source. Creating merged spectra is particularly bad, because it involves merging data taken from different parts of the detector, and at different off-axis angles, making the ancillary response extremely complicated and unreliable. However, simultaneous fitting of the individual observations is also unreliable for sources that vary in luminosity or spectral shape, especially if those observations have poor statistics.

\subsubsection{HRC}

For HRC observations, we we included only PI channels 48--293,  thereby reducing the instrumental background. We used the WebPIMMS tool to find the unabsorbed luminosity equivalent to 1 count s$^{-1}$ on-axis, assuming the same emission model as for the ACIS observations with $<$200 photons.  We created a 1 keV exposure map for each observation, and compared the exposure within the source region with that of an equivalent on-axis region, in order to estimate the necessary exposure. We multiplied the background subtracted, corrected source intensity by the resulting  conversion factor  to get the 0.3--10 keV luminosity.

\subsubsection{Creating the lightcurves}

We created long term 0.3--10 keV  lightcurves for each source, using the luminosities obtained from each observation as described above. We only included observations with net source counts $\ge$ 0 after background subtraction. We fitted each long term lightcurve with a line of constant intensity, in order to ascertain source variability.

We also made 0.5--4.5 keV flux lightcurves for each source that allowed us to make SFs that were directly comparable with the ensemble AGN SF constructed by Vagnetti et al. (2011). We calculated the 0.5--4.5 keV fluxes using the best fit emission model where possible; if no spectral fits were made for a source, then we assumed a typical spectrum with  $N_{\rm H}$ = 7$\times 10^{20}$ atom cm$^{-2}$, and a power law with spectral index 1.7. 

We note that the structure function for each source is expected to be rather insensitive to the emission model used to convert from intensity to flux. This conversion is a scaling factor that is applied to the intensity lightcurve after correcting for instrumental effects; hence, the emission model is only important if the parameters significantly change between observations. 

\subsection{Characteristics  of the structure function}
In this work, we compare the SFs of our X-ray sources with the ensemble AGN SF derived from the XMM-Newton data by \citet{vagnetti11}: SF($\tau$) $\propto$ $\tau^{0.10\pm0.01}$. We estimated the normalization of this relation to be 0.11 from Figure 5 of \citet{vagnetti11}. Therefore the 3$\sigma$ upper limit to the AGN SF derived from the XMM-Newton data is given by 0.11$\tau^{0.13}$.

 \citet{vagnetti11} calculated the noise component from
\begin{equation}
\sigma_n^2 = 2\left<\left(\sigma \log f_{\rm X}\right)^2\right> \simeq 2\left(\log e\right)^2 \left< \left( \frac{\sigma f_{\rm X}}{f_{\rm X}}\right)^2\right>,
\end{equation}
assuming that $\sigma f_{\rm X}/f_{\rm X}$ = $\left(1/N_{\rm phot}\right)^{0.5}$, and $N_{\rm phot}$ is the number of photons. This assumes that all of the noise in the SF comes from Poisson statistics.

Our lightcurves are background subtracted, and ARF-corrected; furthermore, uncertainties in the luminosities of bright sources include uncertainties in the spectral parameters.  As a result, our uncertainties are not simply due to photon counting noise, and may be considerably larger. Hence in our case,
\begin{equation}
\sigma_n^2 \simeq \left(\log e\right)^2 \left< \left[ \frac{\sigma f_{\rm X}\left(t+\tau\right)}{f_{\rm X}\left(t+\tau\right)}\right]^2 + \left[ \frac{\sigma f_{\rm X}\left(t\right)}{f_{\rm X}\left(t\right)} \right]^2\right>,
\end{equation}
where $\sigma$ $f_{\rm X}$ is uncertainty in X-ray flux of a particular observation. The simplest possible SF comes from a single pair of observations with X-ray fluxes $f_1$ and $f_2$; in this case, 
\begin{equation}
\sigma_n^2 \simeq \left(\log e\right)^2 \left[ \left( \frac{\sigma f_1}{f_1}\right)^2 + \left( \frac{\sigma f_2}{f_2} \right)^2\right].
\end{equation}

A constant X-ray source will yield a SF consistent with zero in all $\tau$ channels. However, if the variation is smaller than the uncertainty in a given channel, then the SF is imaginary for that channel. Our SFs show imaginary channels with zero power but finite uncertainties; channels that contain no observation pairs have zero power and zero uncertainties. 

For each source, we calculated the probability, $P_{\rm AGN}$, that it was consistent with being a typical AGN like the ones studied by \citet{vagnetti11}. This was derived from $E_i$, the excess in channel $i$ over the 3$\sigma$ limit to the AGN SF observed with XMM-Newton (0.11$\tau^{0.13}$); each SF contained 19 $\tau$ channels.
\begin{equation}
P_{\rm AGN} = \prod_{i=1}^{19} p_{ i},
\end{equation}
where $p_i$ = 1 when $E_i$ $\le$ 0, and 
\begin{equation}
p_i = 1-\frac{2}{\sqrt{\pi}} \int_{0}^{E_i/\sqrt{2}}e^{-t^2}dt \equiv {\rm erfc}\left(E_i/\sqrt{2}\right)
\end{equation}
when $E_i$ $>$ 0. We then assigned a Rank to each source, given by $-\log\left( P_{\rm AGN} \right)$. A Rank of 2.6 indicates a 3$\sigma$ excess in SF($\tau$) over 0.11$\tau^{0.13}$, while a Rank of 6.2 indicates a 5$\sigma$ excess.

\section{Results}
\label{res}
\subsection{Overview}


\begin{table*}
\begin{center}
\caption{Summary of our findings for the 528 X-ray sources in our survey. For each object, we provide its position with respect to the LGS M31 Field 5 B band image provided by \citet{massey06}, along with 1$\sigma$ uncertainties in RA and Dec. We then give the number of ACIS and HRC observations, $O_{\rm A }$ and $O_{\rm H}$ respectively. Next we show the known properties--- first, we give the classification assigned by \citet{stiele11}, with any further classification following an equals sign. This is followed by the best fit constant luminosity to the unabsorbed 0.3--10 keV luminosity lightcurve and corresponding $\chi^2$/dof. Finally we   rank the structure function for variability.  } \label{sumtab}
\renewcommand{\arraystretch}{.9}
\begin{tabular}{ccccccccccccccc}
\tableline\tableline
Src & Pos &$\sigma_{\rm RA}$/$"$ &$\sigma_{\rm Dec}$/$"$ &  $O_{\rm A}$ & $O_{\rm H}$& Properties& $L_{\rm con}$ /10$^{37}$ & $\chi^2$/dof & SF Rank\\
\tableline 
1 & 00:41:05.41 +41:18:14.7$^a$ & 1.3$^a$ & 1.3$^a$ & 11 & 8 & H & 0.11$\pm$0.03 &  13/18 & 7.3$^f$\\
2 & 00:41:05.46 +41:17:34.1$^a$ & 1.0$^a$ & 1.0$^a$ & 16 & 11 & H & 0.062$\pm$0.019 &  22/26 & 0.5$^g$\\
3 & 00:41:07.92 +41:13:44.3$^a$ & 0.8$^a$ & 0.8$^a$ & 21 & 12 & H & 0.084$\pm$0.019$^d$ &  36/32 & 0.9$^g$\\
4 & 00:41:09.15 +41:15:17.9$^a$ & 1.2$^a$ & 1.2$^a$ & 20 & 17 & H & 0.063$\pm$0.016 &  16/36 & 0.2$^g$\\
5 & 00:41:10.73 +41:21:20.0$^a$ & 0.9$^a$ & 0.9$^a$ & 12 & 20 & [*]=* & 0.11$\pm$0.02 &  219/31 & 51.5\\
6 & 00:41:11.40 +41:20:05.6$^a$ & 0.9$^a$ & 0.9$^a$ & 13 & 12 & [*] & 0.09$\pm$0.02$^d$ &  8/24 & 0.0\\
7 & 00:41:11.83 +41:23:43.6$^a$ & 1.4$^a$ & 1.4$^a$ & 9 & 9 & H & 0.10$\pm$0.03 &  10/17 & 0.0$^g$\\
8 & 00:41:12.42 +41:14:57.6$^a$ & 0.8$^a$ & 0.8$^a$ & 19 & 25 & H & 0.139$\pm$0.017 &  59/43 & 2.6$^f$\\
9 & 00:41:13.96 +41:14:34.5$^a$ & 1.2$^a$ & 1.2$^a$ & 22 & 26 & H & 0.105$\pm$0.016$^d$ &  55/47 & 1.4$^g$\\
10 & 00:41:14.48 +41:14:00.2$^a$ & 1.1$^a$ & 1.1$^a$ & 20 & 27 & [*] & 0.081$\pm$0.015 &  43/46 & 0.5\\
11 & 00:41:15.77 +41:23:28.5$^a$ & 1.9$^a$ & 1.9$^a$ & 10 & 17 & [*]=* & 0.08$\pm$0.02$^e$ &  208/26 & 7.3\\
12 & 00:41:17.49 +41:17:26.3$^a$ & 0.9$^a$ & 0.9$^a$ & 22 & 26 & H & 0.078$\pm$0.016$^d$ &  34/47 & 0.2$^g$\\
13 & 00:41:17.74 +41:12:55.9$^a$ & 1.0$^a$ & 1.0$^a$ & 24 & 18 & H & 0.077$\pm$0.015 &  23/41 & 1.0$^g$\\
14 & 00:41:18.09 +41:10:21.7$^a$ & 1.0$^a$ & 1.0$^a$ & 22 & 16 & H & 0.081$\pm$0.018$^c$ &  11/21 & 0.0$^g$\\
15 & 00:41:18.92 +41:08:28.0   & 1.2 & 1.1 & 1 & 9 & H & 0.08$\pm$0.04 &  9/9 & 4.7$^f$\\
16 & 00:41:18.98 +41:09:07.6$^a$ & 1.3$^a$ & 1.3$^a$ & 12 & 6 & H & 0.06$\pm$0.02 &  14/17 & 0.0$^g$\\
17 & 00:41:20.99 +41:16:01.2$^a$ & 1.3$^a$ & 1.3$^a$ & 19 & 31 & H & 0.074$\pm$0.015$^d$ &  21/49 & 0.0$^g$\\
18 & 00:41:21.70 +41:07:55.9$^b$ & 0.8 & 0.9 & 20 & 4 & NC & 0.066$\pm$0.014 &  10/23 & 0.0$^g$\\
19 & 00:41:22.62 +41:21:35.3$^a$ & 1.0$^a$ & 1.0$^a$ & 18 & 45 & H & 0.139$\pm$0.019$^c$ &  12/17 & 0.0$^g$\\
20 & 00:41:23.25 +41:28:56.9$^a$ & 1.0$^a$ & 1.0$^a$ & 4 & 8 & H  & 0.16$\pm$0.05 &  8/11 & 0.1$^g$\\
21 & 00:41:23.98 +41:15:09.5$^a$ & 0.9$^a$ & 0.9$^a$ & 20 & 32 & [*] & 0.095$\pm$0.016$^d$ &  37/51 & 0.0\\
22 & 00:41:24.9 +41:17:09$^a$ & 3$^a$ & 3$^a$ & 22 & 33 & NC & 0.084$\pm$0.015$^d$ &  45/54 & 0.0$^g$\\
23 & 00:41:25.18 +41:19:44.6$^a$ & 1.1$^a$ & 1.1$^a$ & 16 & 35 & * & 0.080$\pm$0.018 &  43/50 & 0.0\\
24 & 00:41:25.91 +41:12:59.3$^a$ & 1.2$^a$ & 1.2$^a$ & 19 & 30 & H & 0.084$\pm$0.016$^d$ &  34/48 & 0.0$^g$\\
25 & 00:41:27.22 +41:20:03.3$^a$ & 1.1$^a$ & 1.1$^a$ & 18 & 39 & H & 0.072$\pm$0.015$^d$ &  39/56 & 0.4$^g$\\
26 & 00:41:29.3 +41:10:41$^a$ & 2$^a$ & 2$^a$ & 18 & 32 & [*] & 0.080$\pm$0.018$^d$ &  59/49 & 0.0\\
27 & 00:41:29.72 +41:27:44.4$^a$ & 1.9$^a$ & 1.9$^a$ & 8 & 8 & [*] & 0.07$\pm$0.02$^d$ &  5/15 & 1.0\\
28 & 00:41:30.28 +41:28:37.7$^a$ & 1.1$^a$ & 1.1$^a$ & 6 & 10 & [*] & 0.05$\pm$0.03$^d$ &  19/15 & 0.0\\
29 & 00:41:31.17 +41:05:03.0 & 0.8 & 0.9 & 16 & 14 & NE & 0.058$\pm$0.012 &  173/29 & 0.0$^g$\\
30 & 00:41:31.72 +41:15:21.8$^a$ & 1.1$^a$ & 1.1$^a$ & 26 & 39 & H & 0.118$\pm$0.016$^d$ &  53/64 & 0.5$^g$\\
31 & 00:41:31.95 +41:17:46.3$^a$ & 0.7$^a$ & 0.7$^a$ & 28 & 42 & H & 0.30$\pm$0.02 &  69/69 & 3.4$^f$\\
32 & 00:41:34.25 +41:17:45.2$^a$ & 1.6$^a$ & 1.6$^a$ & 24 & 34 & H & 0.112$\pm$0.015 &  52/57 & 0.0$^g$\\
33 & 00:41:35.74 +41:06:56.3 & 0.8 & 1.0 & 25 & 20 & SNR & 0.066$\pm$0.010 &  34/44 & 0.0\\
34 & 00:41:36.08 +41:17:43.0$^a$ & 1.7$^a$ & 1.7$^a$ & 20 & 37 & H=[*],* & 0.110$\pm$0.013$^d$ &  39/56 & 0.0\\
35 & 00:41:36.24 +41:12:24.4$^a$ & 1.5$^a$ & 1.5$^a$ & 26 & 35 & H & 0.086$\pm$0.017 &  46/60 & 0.0$^g$\\
36 & 00:41:40.13 +41:10:34.9$^a$ & 1.6$^a$ & 1.6$^a$ & 18 & 35 & H & 0.106$\pm$0.017$^d$ &  42/52 & 0.0$^g$\\
37 & 00:41:40.15 +41:02:46.9 & 1.5 & 1.3 & 17 & 11 & NE=X & 0.038$\pm$0.010 &  16/27 & 0.1$^g$\\
38 & 00:41:41.30 +41:03:33.1 & 0.8 & 1.1 & 26 & 13 & [AGN] & 0.118$\pm$0.015 &  43/38 & 3.5$^f$\\
39 & 00:41:41.45 +41:19:15.6$^a$ & 0.7$^a$ & 0.7$^a$ & 19 & 46 & !AGN & 0.42$\pm$0.02$^d$ &  93/64 & 10.4$^f$\\
40 & 00:41:41.69 +41:19:54.1$^a$ & 0.9$^a$ & 0.9$^a$ & 16 & 33 & H & 0.162$\pm$0.015 &  36/48 & 0.0$^g$\\
41 & 00:41:43.48 +41:05:04.7 & 0.9 & 0.8 & 21 & 47 & * & 0.079$\pm$0.012$^e$ &  552/67 & 16.8\\
42 & 00:41:43.51 +41:21:18.2$^a$ & 0.6$^a$ & 0.6$^a$ & 9 & 32 & [XB] & 0.112$\pm$0.019 &  46/40 & 3.7$^f$\\
43 & 00:41:44.72 +41:11:09.5 & 0.5 & 0.6 & 30 & 54 & NE & 2.04$\pm$0.06 &  2914/83 & 43.6$^f$\\
44 & 00:41:45.34 +41:26:24.1$^a$ & 0.7$^a$ & 0.7$^a$ & 16 & 56 & [AGN] & 0.352$\pm$0.018 &  368/71 & 11.2$^f$\\
45 & 00:41:46.76 +41:16:56$^a$ & 2$^a$ & 2$^a$ & 25 & 36 & [AGN] & 0.074$\pm$0.013$^d$ &  44/60 & 0.0$^g$\\
46 & 00:41:46.96 +41:23:16.7$^a$ & 1.5$^a$ & 1.5$^a$ & 15 & 39 & H & 0.080$\pm$0.014 &  50/53 & 0.4$^g$\\
47 & 00:41:47.55 +41:14:24.9$^a$ & 1.7$^a$ & 1.7$^a$ & 16 & 34 & H & 0.089$\pm$0.015$^d$ &  22/49 & 0.0$^g$\\
48 & 00:41:48.23 +41:07:07.9 & 1.0 & 1.0 & 22 & 31 & H & 0.060$\pm$0.011 &  29/52 & 2.2$^g$\\
49 & 00:41:49.60 +41:03:33.5 & 1.0 & 0.8 & 14 & 16 & NE & 0.08$\pm$0.03 &  20/29 & 0.0$^g$\\
50 & 00:41:49.75 +41:01:07.4$^{b}$ & 1.9 & 1.7 & 21 & 17 & H & 0.65$\pm$0.03 &  396/37 & 65.3$^f$\\
51 & 00:41:50.28 +41:13:35.7 & 0.8 & 0.6 & 20 & 53 & [AGN] & 0.272$\pm$0.016$^d$ &  86/72 & 1.2$^g$\\
52 & 00:41:50.43 +41:12:10.4 & 0.9 & 1.0 & 28 & 38 & GC & 0.166$\pm$0.014 &  53/65 & 0.0$^f$\\
53 & 00:41:50.71 +41:21:14.3 & 0.5 & 0.8 & 24 & 47 & H & 0.163$\pm$0.011 &  54/70 & 0.0$^g$\\
54 & 00:41:51.00 +41:06:48.9$^a$ & 1.4$^a$ & 1.4$^a$ & 21 & 41 & H & 0.099$\pm$0.016 &  52/61 & 0.3$^g$\\
55 & 00:41:51.66 +41:14:39.1 & 0.4 & 0.4 & 29 & 58 & !AGN & 0.54$\pm$0.02 &  121/86 & 6.8$^f$\\
56 & 00:41:51.86 +41:15:17.1$^a$ & 1.6$^a$ & 1.6$^a$ & 27 & 33 & [*] & 0.073$\pm$0.013 &  51/59 & 2.1\\
57 & 00:41:54.31 +41:07:23.2$^a$ & 0.7$^a$ & 0.7$^a$ & 18 & 38 & [SSS]=[Nova] & 0.083$\pm$0.017$^d$ &  35/55 & 0.0\\
58 & 00:41:54.83 +41:07:34.3 & 0.6 & 1.1 & 20 & 32 & H & 0.134$\pm$0.019 &  38/51 & 0.0$^g$\\
59 & 00:41:55.24 +41:23:01.8 & 0.8 & 0.8 & 16 & 48 & H & 0.105$\pm$0.009$^d$ &  107/63 & 27.1$^f$\\
60 & 00:41:57.24 +41:14:40.2$^a$ & 1.5$^a$ & 1.5$^a$ & 41 & 29 & H & 0.053$\pm$0.011$^d$ &  38/69 & 0.0$^g$\\
\tableline
\end{tabular}
\\$^a$ Positions and uncertainties obtained from \citet{stiele11}; $^b$ position uncertainty obtained from a 9$\times$9 binned image; $^c$ HRC observations ignored; $^d$ HRC observations included in luminosity fit but not SF; $^e$ fitted with blackbody model $N_{\rm H}$ =7$\times 10^{20}$ atom cm$^{-2}$, k$T$ = 0.05 keV; $^f$ likely XB; $^g$ consistent with AGN.
\end{center}
\end{table*}


We will first provide an overview of our results before dealing with specific aspects of our data. Table~\ref{sumtab} summarizes our main results. For each source, we give its position with respect to the LGS M31 Field 5 B band image provided by \citet{massey06}, along with 1$\sigma$ position uncertainties. Sources marked with $^a$ have locations derived from XMM-Newton observations, and 1$\sigma$ uncertainties calculated from the 3$\sigma$ uncertaintites quoted by \citet{stiele11}. Sources marked with $^b$ have positions derived from a highly binned image, where one image pixel is equivalent to 9$\times$9 native pixels. We then show the number of ACIS and HRC observations that contain the source. 

Next in Table~\ref{sumtab} we present the published source classification and properties. We first show the classification in the \citet{stiele11} XMM-Newton catalog of M31 X-ray sources; sources can be classified as XBs (XB), globular clusters (GC), stars (*), galaxies (Gal), AGN (AGN), supernova remnants (SNR), supersoft sources (SSS), or simply hard (H); square brackets indicate candidates, while exclaimation marks  indicate classifications that were rejected for our Chandra locations. The remaining X-ray sources either were observed but not classified (NC), or had no entry at all (NE). Any improvement on the \citet{stiele11} classification is indicated by "="; these can be known X-ray sources that do not appear in the XMM-Newton catalog (X), black hole candidates (BHC), transients (T), novae (Nova), ultra-luminous X-ray sources \citep[U, exhibiting X-ray luminosities $>$2$\times 10^{39}$ erg s$^{-1}$][]{kaur12,nooraee12,middleton13,barnard13c}, or variable stars (var*).

Finally in Table~\ref{sumtab} we show the best fit  constant 0.3--10 keV luminosity over our $\sim$13 year monitoring program, along with the corresponding $\chi^2$/dof, and our ranking of the object according to the variablity shown in its structure function. Sources with Rank $>$2.6 are significantly more variable than expected for typical AGN. For 28 sources indicated with $^c$, we ignored the HRC observations completely; for 120 sources indicated with $^d$, we included HRC observations in the luminosity calculation but exclude HRC data from the SF; reasons for these decisions are given below. The  10 sources indicated with $^e$ are very soft, and were modelled using an absorbed blackbody with $N_{\rm H}$ = 7$\times 10^{20}$ atom cm$^{-2}$ and k$T$ = 0.05 keV; using our standard emisson model for these sources resulted in large  systematic offsets between ACIS and HRC luminosities. Likely X-ray binaries are indicated by $^f$, while $^g$ indicates sources that are consistent with AGN. 

Our luminosities assume a distance of 780 kpc \citep{stanek98}; clearly the distances to foreground stars and background AGN will be very different, along with their corresponding luminosities. In each case, the flux is given by the luminosity divided by 7.3$\times 10^{49}$ cm$^{2}$.

The following sections describe our source list, spectral analysis, and time variability studies in detail.

\subsection{ The source list}
We initially identified 407 X-ray sources from our Chandra observations; registering the merged 0.3--7 keV image to the LGS B Band optical image resulted in 1$\sigma$ positional uncertainties of 0.11$''$ in RA and 0.09$''$ in Dec. Each X-ray source also has an uncertainty in its location on the X-ray image, which is dependent on the source intensity and off-axis angle.

We compared our X-ray sources with the M31 XMM-Newton source catalog of \citet{stiele11}, by looking for matches within 10$''$ of our Chandra positions; these associations were then accepted or rejected based on the 3$\sigma$ uncertainties in XMM-Newton position reported by \citet{stiele11}. The XMM-Newton catalog of \citet{stiele11} contained $\sim$170 X-ray sources that were not in our catalog. We extracted lightcurves and spectra from regions at the positions of these sources, and found evidence for a further 121 X-ray sources; we found no trace of $\sim$50 of the X-ray sources in the \citet{stiele11}. We obtained the positions and uncertainties for  the 121 new X-ray sources from \citet{stiele11}; our total source list contains 528 X-ray sources.

We found that 208 of our X-ray sources had no counterpart in the XMM-Newton catalog despite both catalogs having similar   luminosity limits  ($\ga$10$^{35}$ erg s$^{-1}$).  While 36 of these sources are transient, the remainder were likely missed by XMM-Newton because of its larger PSF and higher background.  
 A further 169 X-ray sources were classified only as ``Hard'' in \citet{stiele11}, while another 9 X-ray sources were detected but not classified. Identified sources included  38 GCs or GC candidates; 15 AGN or candidates, including 2 optically identified AGN; 4 novae or candidates; 13 supernova remnants or  candidates; 28 foreground stars or candidates; 21 SSS candidates; and 27 XBs or XB candidates including 5 BHCs and 13 transients \citep{stiele11}. 

We rejected 4 AGN, 2 GC and 3 SN candidates identified in the \citet{stiele11} after obtaining our superior Chandra positions.
Our total inventory includes 47 foreground stars or candidates; 7 novae or nova  candidates; 13 supernova remnants or candidates; and 13 supersoft source candidates with no further classification. Also included are 47 GCs or GC candidates; 36 black hole candidates (12 in GCs), with 1 GC BHC new to this work; and 52  transients (6 located in GCs).

We estimated the probable number of false associations by finding the number of matches between our X-ray sources and astronomical objects after  adding 20$''$ to  or subtracting 20$''$ from the Right Ascension or Declination. In each of the four tests we found 6--11 false matches with stars, 0--1 false matches with a candidate supernova remnant, 0--2 false matches with candidate globular clusters, and 1--4 false matches with novae.

\subsection{Spectral fitting}

We found 120 sources that had  spectra with $>$200 net source counts in at least one  ACIS observation. We fitted the spectra with the emission model that suited the source classification; stars were fitted with blackbody models, X-ray transients in outburst were fitted with disk blackbody models \citep[see e.g.][]{barnard2012b}, and the remaining sources were fitted with power law models; each model included line-of-sight absorption, with a minimum value equivalent to 7$\times 10^{20}$ H atom cm$^{-2}$.

 High quality XMM-Newton observations of the brightest sources in our survey required two-component models \citep[e.g. blackbody   + power law, see][ for an example]{barnard06}; however, Chandra spectra from these sources were acceptably fitted by simple power laws ($\Gamma$ $\la$ 1). Due to the difference between HRC and ACIS responses, we ignored the HRC data for these systems; even if we used two component emission models, it would be impossible to estimate the HRC luminosities due to the uncertainties in the  contributions of the different components.


\begin{table*}
\begin{center}
\caption{ Spectral fits for sources with $>$200 net source counts in at least one ACIS observation. } \label{spectab}
\renewcommand{\arraystretch}{.9}
\begin{tabular}{ccccccccccccccc}
\tableline\tableline
Src   &   Model   &   $N_{\rm H}$/10$^{22}$   &   $\chi^2$/dof   &   Param   &  $\chi^2$/dof \\
\tableline 
31 & PO & 0.3$\pm$0.2 & --- & 2.3$\pm$0.9 & ---\\
43 & PO & 0.31$\pm$0.04 & 4/11 & 2.37$\pm$0.11 & 3/11\\
55 & PO & 0.19$\pm$0.12 & --- & 2.1$\pm$0.4 & ---\\
78 & PO & 0.34$\pm$0.04 & 8/9 & 1.85$\pm$0.07 & 3/9\\
83 & PO & 0.33$\pm$0.14 & 2/3 & 1.57$\pm$0.18 & 1.1/3\\
84 & PO & 0.50$\pm$0.13 & 1.6/5 & 2.3$\pm$0.2 & 1.4/5\\
90 & PO & 0.11$\pm$0.07 & 0.13/2 & 1.7$\pm$0.2 & 0.31/2\\
95 & PO & 0.45$\pm$0.09 & 0.2/3 & 1.85$\pm$0.13 & 1.4/3\\
100 & BB & 0.129$\pm$0.019 & 12/47 & 0.572$\pm$0.007 & 37/47\\
103 & PO & 0.28$\pm$0.06 & 0.13/4 & 2.12$\pm$0.13 & 4/4\\
106 & PO & 0.12$\pm$0.12 & --- & 1.7$\pm$0.4 & ---\\
108 & PO & 0.41$\pm$0.09 & 1.1/2 & 2.05$\pm$0.18 & 7/2\\
109 & PO & 0.5$\pm$0.06 & 5/17 & 1.51$\pm$0.08 & 5/17\\
111 & PO & 0.6$\pm$0.2 & 0.06/1 & 1.9$\pm$0.3 & 0.3/1\\
117 & DBB & 0.06$\pm$0.04 & 0.006/1 & 0.38$\pm$0.02 & 14/1\\
120 & PO & 0.30$\pm$0.04 & 8/24 & 2.88$\pm$0.12 & 10/24\\
122 & PO & 0.16$\pm$0.04 & 5/15 & 1.47$\pm$0.07 & 4/15\\
142 & PO & 0.41$\pm$0.04 & 9/14 & 2.16$\pm$0.07 & 8/14\\
143 & PO & 0.18$\pm$0.09 & 0.3/1 & 2.2$\pm$0.2 & 0.002/1\\
146 & PO & 0.13$\pm$0.13 & --- & 1.2$\pm$0.4 & ---\\
148 & PO & 0.14$\pm$0.04 & 4/5 & 1.74$\pm$0.09 & 6/5\\
151 & PO & 0.125$\pm$0.009 & 77/86 & 1.552$\pm$0.015 & 76/86\\
153 & PO & 0.13$\pm$0.05 & 1.9/6 & 1.79$\pm$0.13 & 1.9/6\\
158 & PO & 0.18$\pm$0.06 & 1.2/6 & 2.2$\pm$0.1 & 4/6\\
159 & PO & 0.43$\pm$0.05 & 1.5/9 & 1.89$\pm$0.07 & 6/9\\
160 & PO & 0.26$\pm$0.12 & 0.7/4 & 1.7$\pm$0.2 & 0.4/4\\
167 & PO & 0.16$\pm$0.04 & 4/5 & 1.5$\pm$0.1 & 6/5\\
168 & PO & 0.11$\pm$0.02 & 11/37 & 1.58$\pm$0.03 & 14/37\\
171 & PO & 0.1$\pm$0.05 & 2/5 & 2.05$\pm$0.12 & 3/5\\
175 & BB & 0.07$\pm$f & --- & 0.25$\pm$0.03 & ---\\
179 & PO & 0.159$\pm$0.012 & 39/94 & 1.69$\pm$0.03 & 49/94\\
181 & PO & 0.25$\pm$0.04 & 3/8 & 1.90$\pm$0.09 & 3/8\\
184 & PO & 0.068$\pm$0.013 & 11/38 & 1.25$\pm$0.03 & 23/38\\
188 & PO & 0.18$\pm$0.07 & 1.4/6 & 1.65$\pm$0.14 & 1.1/6\\
195 & PO & 0.08$\pm$0.03 & 1.5/8 & 1.61$\pm$0.07 & 3/8\\
198 & PO & 0.15$\pm$0.07 & 1.5/4 & 1.37$\pm$0.15 & 2/4\\
201 & PO & 0.31$\pm$0.10 & 4/9 & 1.72$\pm$0.17 & 3/8\\
209 & PO & 0.073$\pm$0.007 & 45/104 & 0.856$\pm$0.008 & 370/104\\
213 & PO & 0.22$\pm$0.05 & 2/5 & 2.36$\pm$0.16 & 3/5\\
214 & PO & 0.1$\pm$0.02 & --- & 1.74$\pm$0.09 & ---\\
217 & PO & 0.113$\pm$0.014 & 8/16 & 1.88$\pm$0.04 & 22/16\\
219 & PO & 0.26$\pm$0.05 & 3/9 & 1.73$\pm$0.08 & 4/9\\
223 & PO & 0.1$\pm$0.02 & 6/11 & 1.64$\pm$0.05 & 10/11\\
229 & PO & 0.11$\pm$0.05 & 1.1/5 & 1.73$\pm$0.14 & 0.3/5\\
231 & PO & 0.3$\pm$0.30 & --- & 1.5$\pm$0.5 & ---\\
233 & PO & 0.3$\pm$0.2 & 0.05/1 & 2.1$\pm$0.4 & 0.5/1\\
236 & PO & 0.09$\pm$0.03 & 1.1/9 & 1.41$\pm$0.07 & 4/9\\
238 & PO & 0.07$\pm$0.02 & 3/10 & 1.95$\pm$0.06 & 4/9\\
241 & PO & 0.12$\pm$0.02 & 5/11 & 1.44$\pm$0.05 & 6/11\\
247 & PO & 0.07$\pm$0.02 & 1.4/10 & 1.52$\pm$0.05 & 6/10\\
249 & PO & 0.5$\pm$0.3 & --- & 2.6$\pm$0.6 & ---\\
250 & BB & 0.07$\pm$f  & --- & 0.21$\pm$0.02 & ---\\
251 & DBB & 0.319$\pm$0.010 & 3/6 & 0.656$\pm$0.007 & 47/6\\
252 & PO & 0.15$\pm$0.02 & 5/10 & 1.75$\pm$0.05 & 8/10\\
253 & PO & 0.12$\pm$0.09 & 1.8/1 & 2.4$\pm$0.3 & 0.16/2\\
254 & PO & 0.20$\pm$0.05 & 3/8 & 1.78$\pm$0.11 & 3/8\\
255 & PO & 0.13$\pm$0.04 & 3/8 & 1.51$\pm$0.08 & 3/8\\
256 & PO & 0.10$\pm$0.04 & 1.6/6 & 1.52$\pm$0.11 & 3/6\\
260 & PO & 0.09$\pm$0.04 & 0.2/4 & 2.12$\pm$0.14 & 2/4\\
264 & PO &0.08$\pm$0.03 &0.8/7 & 2.27$\pm$0.12 & 3/7\\

\tableline
\end{tabular}

\end{center}
\end{table*}


 For sources with multiple spectral fits, we obtained the best fit constant values for the absorption and emission parameter,  along with the corresponding $\chi^2$/dof, where the dof  is one less than the number of fitted observations. Table~\ref{spectab} summarizes our findings.  For each source we provide the emission model used (power law ``PO'', blackbody ``BB'', or disk blackbody ``DBB''), absorption ($N_{\rm H}$/10$^{22}$ atom cm$^{-2}$) plus $\chi^2$/dof, and parameter (photon index for PO, k$T$/keV for BB or DBB) plus $\chi^2$/dof. If only one observation was fitted, then the $\chi^2$/dof is given as ---; in cases with only 2 observations (1 degree of freedom), $\chi^2$ was often extremely low, and meaningless.  The absorption was fixed to Galactic line of sight absorption (7$\times 10^{20}$ atom cm$^{-2}$) for three sources (S175, S250, and S287).

In many cases, the parameter fits have low $\chi^2$/dof; this is to be expected when each datum is itself  derived from fitting a spectrum which may have large uncertainties.  Only one source (S327) exhibited significant variation in absorption. Five sources (S117, S209, S251, S327, and S396) exhibited significant variability in their emission parameters; indeed the variability in S209 was so extreme that the HRC observations were unusable (each ACIS observation had sufficient counts for fitting). S117, S251 and S396 are transients, S209 is a candidate Z-source \citep{barnard03}, and S327 is a black hole candidate \citep{barnard13}.
\begin{figure*}
\epsscale{1.}
\plotone{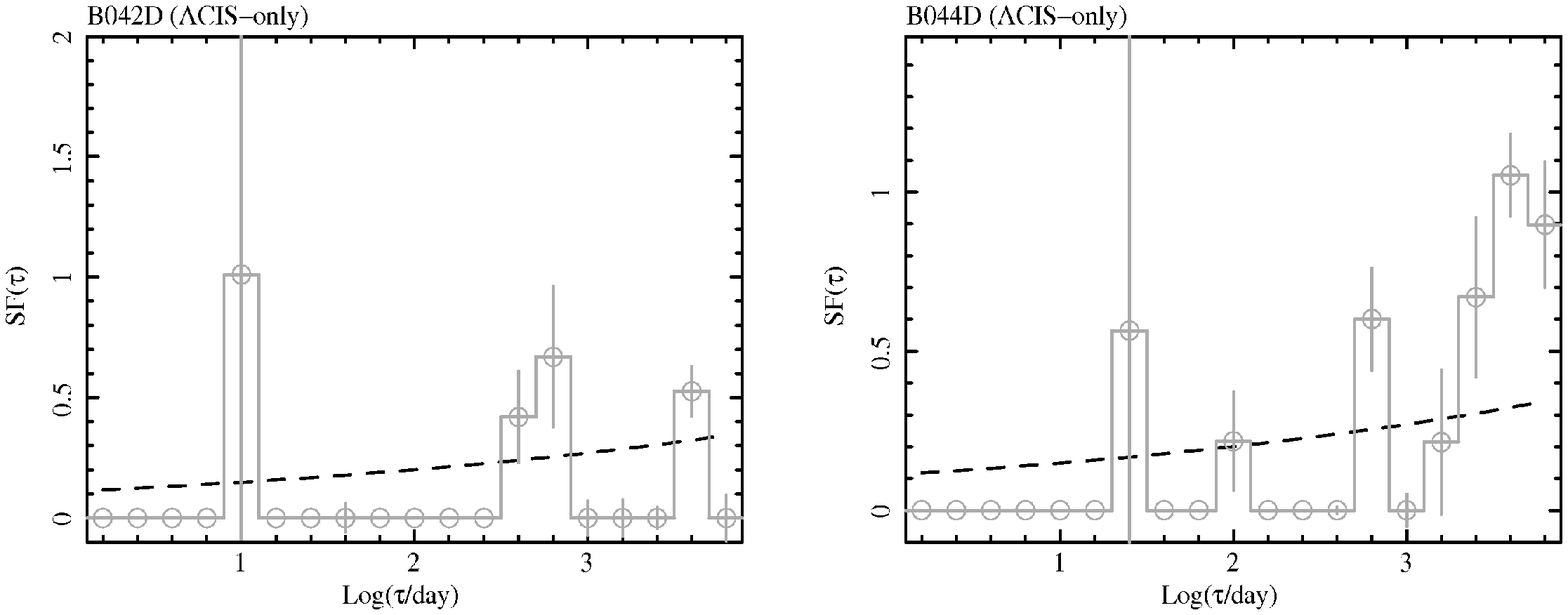}
\caption{0.5--4.5 keV ACIS-only SFs for the two AGN associated with optical galaxies, B042D and B044D.  The dashed curves represent SF($\tau$) = 0.11$\tau^{0.13}$, the 3$\sigma$ upper limit to the ensemble AGN SF created by \citet{vagnetti11} from XMM-Newton data. These galaxies have luminosities $\sim$10$^{42}$ and $\sim 10^{41}$ erg s$^{-1}$ respectively, 1--5 orders of magnitude lower than the galaxies in the sample used by \citet{vagnetti11}. }\label{agnsf}
\end{figure*}
It is unsurprising that so few X-ray sources exhibited spectral variability, since the hard state is observed up to $\sim$0.1 times the Eddington limit ($\sim 2\times 10^{37}$ erg s$^{-1}$ for a 1.4 $M_{\odot}$ NS), and $\sim$90\% of our sample is consistent with being in the NS hard state. For example, S213 has a best fit luminosity $\sim$10$^{37}$ erg s$^{-1}$ and a Rank $>$320, and the entire 0.3--10 keV lightcurve is consistent with  S213 being in a NS hard state. Furthermore, the statistics led to relatively large uncertainties in the emission parameters, making significant variability harder to detect than for Galactic XBs.

\subsection{X-ray identification of  foreground stars}

Many foreground stars, or candidates, were identified by their proximity to an optical counterpart. Some of these stars were identified as Hard by \citep{stiele11}, due to overlapping selection criteria (HR2$-$EHR2 $>$ $-$0.2 for hard sources with no other classifcation, but HR2$-$EHR2 $<$0.3 for foreground stars along with other criteria). We label these sources as ``H=[*]''.

 It was therefore neccessary to investigate these sources further, using the best ACIS spectra available; if the observed counts fell mostly below 1 keV, then we considered the source a foreground star, indicated by ``H=[*],*''. Otherwise we condidered them a ``hard'' source, ``H=[*],H''. If there were too few counts to even determine this much, then we conservatively classify the source as a candidate star, ``H=[*],[*]''. Two sources (S34, and S74) were classified as stars, and four sources (S138, S165, S282, S402) were classified as candidate stars. Seven sources (S87, S92, S340, S399, S405, S499, and S507) were classified as hard.

\subsection{Long-term variability}

We assessed the long-term variability of each X-ray source in two ways: the 0.3--10 keV lightcurves, and the 0.5--4.5 keV SFs. The instrumental responses of the ACIS and HRC instruments are significantly different, and the only the ACIS observations are capable of providing reliable spectral data.
 Therefore, we made two lightcurves and two SFs for each source, one including ACIS and HRC observations, and one including only ACIS data. 

There are two situations in which the inclusion of HRC data may be detremental to our studies of a particular X-ray source. In the first situation, where the HRC lightcurve is systematically offset from the ACIS lightcurve due to differences in instrumental response, we exclude the HRC observations entirely; such cases are indicated by $^c$ in Table~\ref{sumtab}. In the second situation, where particularly large uncertainties in HRC observations dominate the structure functions, we include the HRC observations in our luminosity estimates, but exclude them from our SFs; these cases are indicated by $^d$ in Table~\ref{sumtab}.  Whenever the two SFs for a particular source gave conflicting results, we favored the ACIS-only SF.

The first situation can affect any source, but is particularly important for bright sources exhibiting significant spectral evolution. The ACIS and HRC luminosities only agree if the chosen  emission model is appropriate for the observed spectrum. For most sources, we must assume a standard spectrum; this is successful in many cases, but for those that are unsuccessful, using only the ACIS data minimizes the uncertainties in luminosity and variability.

The second situation affects some faint X-ray sources. Some HRC obsevations have exposure times as short as 1 ks; since the sensitivity of the HRC is a factor $\sim$5 lower than the ACIS for typical X-ray binary spectra, the large uncertainties in  HRC luminosities  can dominate the SF to the extent that variability between ACIS obsevations is lost. 

The 0.5--10 keV flux distribution obtained by \citet{georgakakis08} leads us to expect 635 AGN within our 20$'$ region with fluxes $\ga 1.4\times 10^{-15}$ erg cm$^{-2}$ s$^{-1}$, equivalent to luminosities $> 10^{35}$ erg s$^{-1}$ at 780 kpc; if the sensitivity of our survey were uniform, we would expect to observe $<$2 typical AGN \citep[as defined by][]{vagnetti11} with  Rank $\ge$2.6.  Considering the deterioration of sensitivity with off-axis angle, and the fact that our uncertainties in the background-subtracted luminosities are often substantially larger than the Poisson statistics assumed by \citet{vagnetti11}, we expect $\ll$1 typical AGN to exhibit Rank $\ge$ 2.6.


\begin{figure*}
\epsscale{1}
\plotone{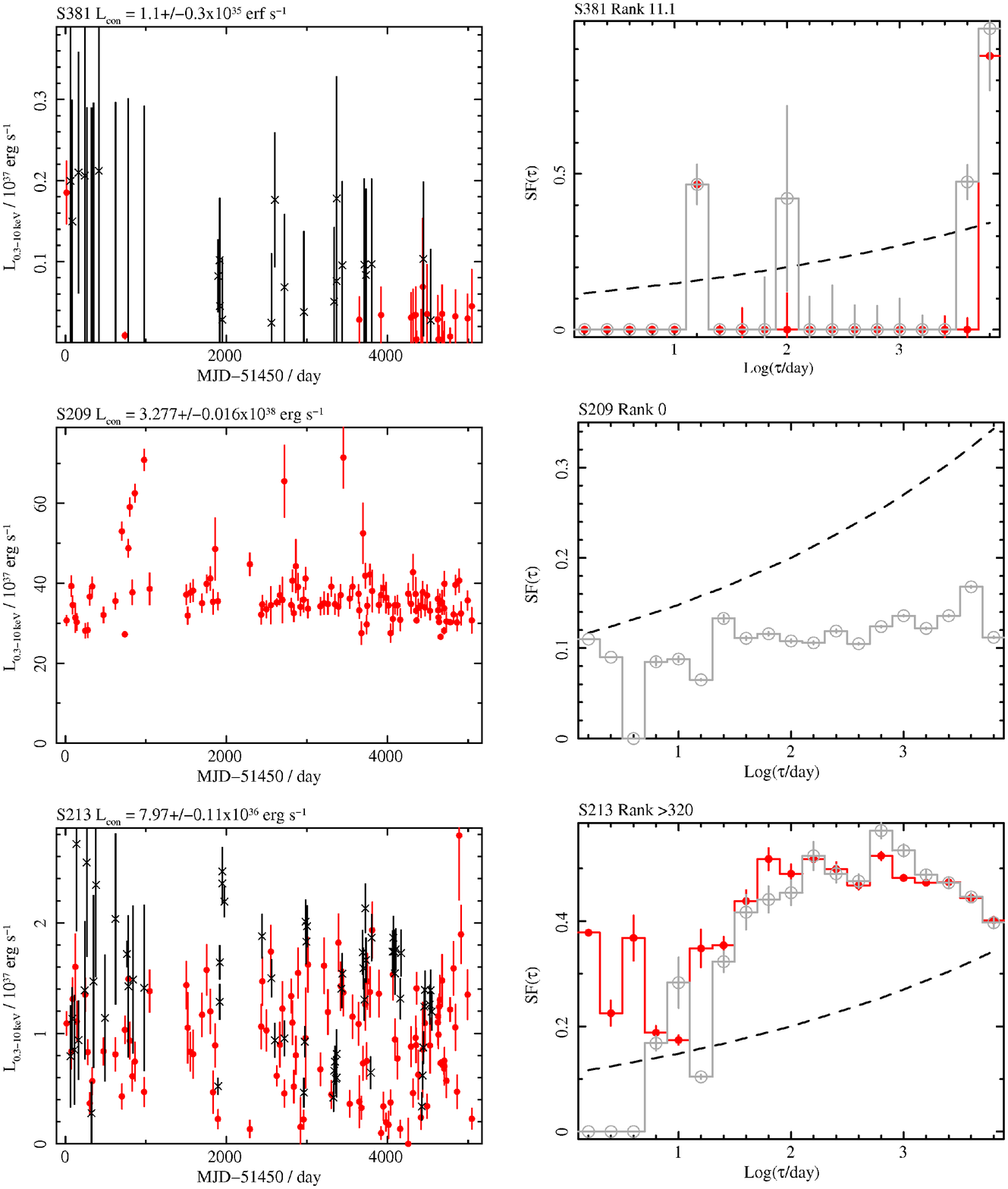}
\caption{Long-term 0.3--10 keV lightcurves and 0.5--4.5 keV structure functions for three representative X-ray sources. S381 is one of the faintest X-ray binary candidates that we identify from its  SF (Rank = 11.1); SF($\tau$) is imaginary for channels where the variation is less than the noise: this is indicated by zero power and finite uncertainties.  S209 is one of our brightest XBs; even though its variability is highly significant, the fractional variation is lower than expected for typical AGN. S213 is representative of our low luminosity XBs with a SF that clearly distinguishes it from an AGN; its Rank is $>$320. In each lightcurve, ACIS and HRC observations are represented by circles and crosses respectively; no HRC data was used for S209.  For the SFs, closed circles represent SFS from the ACIS and HRC observations, while open circles represent the ACIS-only SFs; dashed lines represent SF($\tau$) = 0.11$\tau^{0.13}$. }\label{lcs_sfs}
\end{figure*}

Since we assume a particular emission model to convert from counts to luminosity in many cases, spectral variation in the X-ray sources would lead to systematic uncertainties. However, $\sim$90\% of our sample exhibited 0.3--10 keV luminosities consistent with the NS hard state throughout our 13 year monitoring campaign, hence we expect little spectral variation from these sources. Furthermore, only 5 out of the 120 X-ray sources with freely fitted spectra exhibited significant spectral variation. Hence spectral change is unlikely to be a major source of systematic uncertainties in estimating luminosity variability.

\subsubsection{Classifying the X-ray sources}

 We found 246 sources with SF Rank $\ge$2.6, and 282 sources with Rank $<$2.6.
 Sources with Rank $\ge$2.6 include 18 foreground stars, 5 supersoft sources, 7 supernova remnants, 1 nova, and 2 low-luminosity AGN (discussed below); the supernova remnants are extended, and not expected to be variable, hence we assume that the associations of strongly variable X-ray sources with supernova remnants are coincidental. Sources with Rank $<$2.6 include 11 transients, 13 sources associated with GCs, 29 stars, 8 supersoft sources, 6 novae, and 6 bright X-ray binaries.
\begin{figure}
\epsscale{1.1}
\plotone{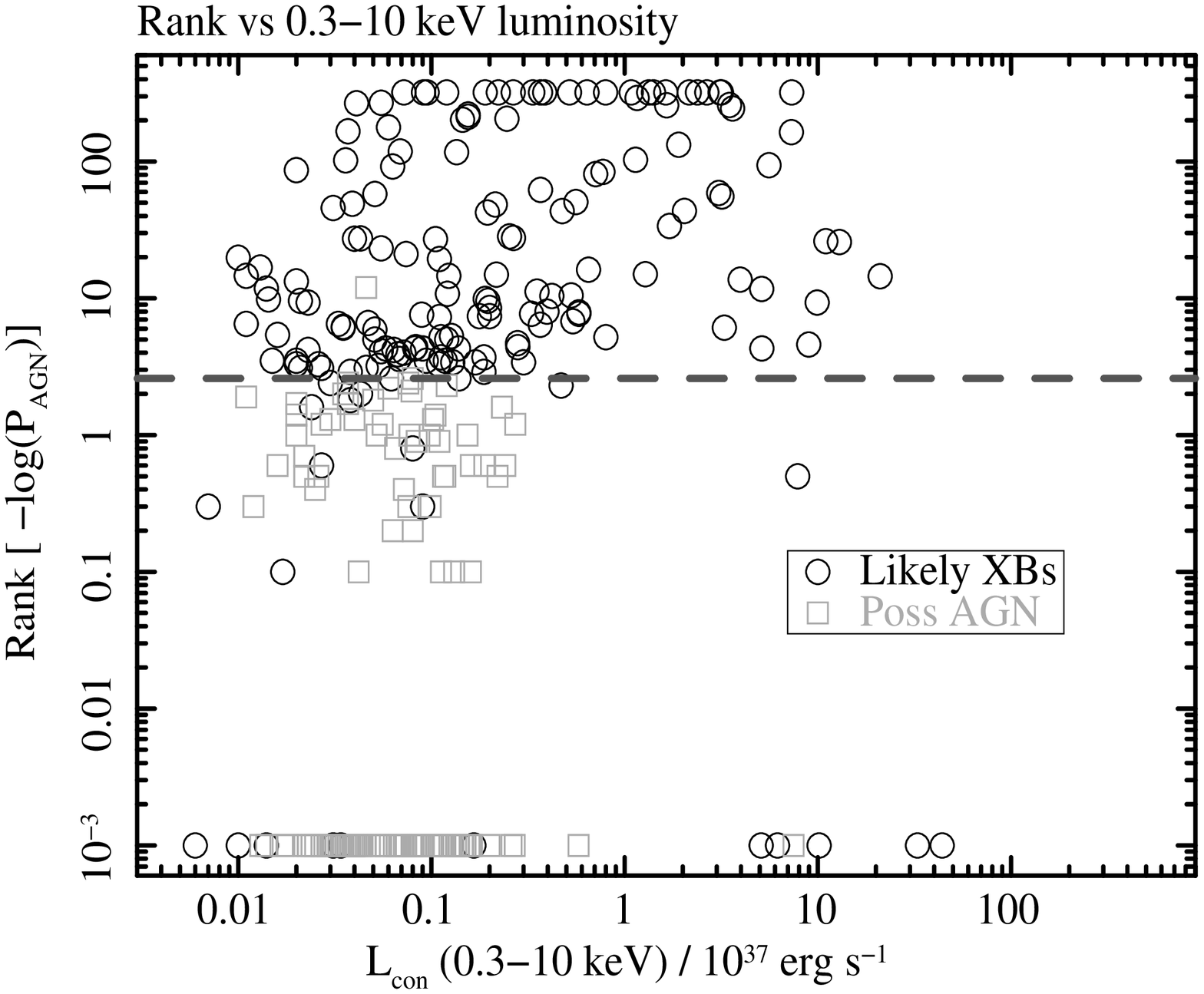}
\caption{Comparison of Rank (i.e. $-$log[$P_{\rm AGN}$]) vs. 0.3--10 keV luminosity for our X-ray binaries (open circles) and sources consistent with AGN (open squares). The dashed line indicates a Rank of 2.6, corresponding to a 3$\sigma$ rejection of an AGN identification. An offset of 0.001 was added to those sources with Rank 0, to allow them to show up on a log scale. }\label{rankvsl}
\end{figure}

 We classify as X-ray binaries all 52 X-ray transients (this number excludes novae), all 47 globular cluster sources,  the 7 variable sources associated with supernova remnants,  144 unclassified X-ray sources with Rank $\ge$2.6, and 6 bright X-ray binaries with Rank $<$2.6; we therefore identify 250 X-ray binaries in total (6 transients are located in GCs).  We do not count the novae or unidentified supersoft sources, although some of them may be XBs. It is perhaps surprising that 11 transients exhibited Ranks $<$2.6; however, the lightcurves for these sources include many observations when they were in quiescence, meaning that the noise dominates the SFs for these sources. We found 161 XBs with Rank $\ge$ 6.4 (a 5$\sigma$ rejection of the AGN classification), and 36 XBs have Ranks $>$320, the limit imposed by machine precision.

We found that 202 unclassified X-ray sources and 1 Galaxy candidate identified by \citet{stiele11} exhibited Ranks $<$2.6, and are consistent with being AGN. In addition there are two optically identified AGN that exhibit high variability: the ACIS-only SF for S73 has Rank 2.6, while the ACIS- only SF for  S75 has Rank 12.3 (shown in Figure~\ref{agnsf}); the ACIS+HRC SFs were dominated by noise from the HRC. However, these galaxies  (B042D and B044D) are relatively  nearby \citep[with velocities of 57833 km s$^{-1}$ and 35743 km s$^{-1}$ respectively,][]{caldwell09}, and exhibited low luminosities ($\sim$10$^{42}$ erg s$^{-1}$ and $\sim$10$^{41}$ erg s$^{-1}$). The ensemble AGN SF created by \citet{vagnetti11} sampled AGN with luminosities 10$^{43}$--$10^{45.5}$ erg s$^{-1}$, $\sim$1--5 orders of magnitude higher than S73 and S75; therefore we estimated the variability seen at 10$^{41}$ and 10$^{42}$ erg s$^{-1}$ using the published relations between luminosity and variability: 
$I_{\rm var}$ $\propto$ $L^{-0.21}$ over $\sim$100 day time-scales \citep{vagnetti11}, and $I_{\rm var}$ $\propto$ $L^{-0.13}$ over year time-scales \citep{markowitz04}. We found that S73 and S75 were both consistent with the expected AGN SF when scaled to their luminosities.  These results do not jeapordize our ranking system, since we expect that  other low-luminosity AGN would have to be local to be detected, and that such nearby AGN would be identified in other wavelengths.

Figure~\ref{lcs_sfs} compares the long-term lightcurves and SFs of three X-ray sources. Panels a) and b) show the lightcurve and SF for S381, one of the faintest X-ray sources with a Rank $>$2.6. S381 has a mean 0.3--10 keV luminosity of 1.1$\pm$0.3$\times 10^{35}$ erg s$^{-1}$, with $\chi^2$/dof = 47/49; its SF yielded a Rank of 11.1.  We note that the SF includes several bins with zero SF but finite uncertainties; SF($\tau$) is imaginary for these bins, because the variation is smaller than the noise. Panels c) and d) give the ACIS-only lightcurve and SF for S209, one of the brightest X-ray sources in the field of view; $\chi^2$/dof = 1255/106 for its best fit constant intensity, yet its SF shows less fractional  variation than a typical AGN over all time-scales. S209 is one of those sources where the HRC data were excluded entirely from the analysis. Panels e) and f) show the lightcurve and SF for S213, one of the 36 XBs with Rank $>$320. These results agree with the well known behavior of Galactic XBs where the high luminosity XBs only vary by a factor of a few, while low luminosity XBs can vary in intensity by 1--2 orders of magnitude \citep[see e.g.][]{muno02}.

In Figure~\ref{rankvsl} we compare the Rank and luminosity for our sources identified as XBs and possible AGN. We see that sources with Rank $<$2.6 are found at 0.3--10 keV luminosities $\la 6\times 10^{36}$ erg s$^{-1}$ and $\ga 5\times 10^{37}$ erg s$^{-1}$; it is well known that high luminosity XBs in our Galaxy are substantially less variable than low luminosity XBs \citep{muno02}. It is possible that the intrinsic variability of some faint XBs was obscured by the scarcity of observed photons. The median Ranks for our XBs, sources consistent with AGN, and the remaining sources are 11.2, 0.0, and 1.0, respectively; clearly our XB population is vastly more variable than the other classes of sources.


\begin{figure}
\epsscale{1}
\plotone{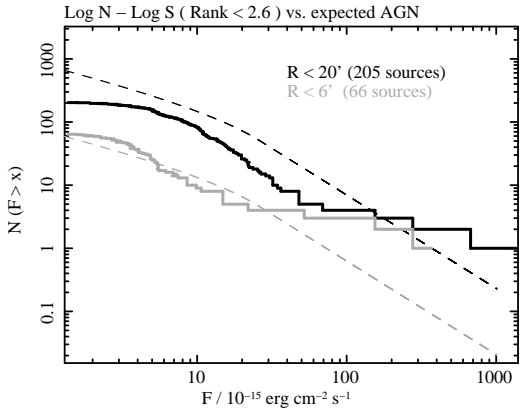}
\caption{Comparison between  the 0.3--10 keV flux distribution of observed AGN candidates (solid lines) and the expected 0.5--10 keV  AGN distribution from \citet{georgakakis08} (dashed lines). The 0.3--10 keV and 0.5--10 keV fluxes agree to within 10\%. Black and gray lines correspond to circular regions with 20$'$ and 6$'$ radius, respectively.  }\label{rank0lf}
\end{figure}

\subsubsection{Comparing our possible AGN  population with expected AGN}

We found 205 X-ray sources to be consistent with AGN; however, this sample may contain some faint XBs.
\citet{georgakakis08} have created a flux distribution (i.e. log N vs log S) for AGN in the 0.5--10 keV band, which is similar enough to our 0.3--10 keV band for comparison; the 0.3--10 keV and 0.5--10 keV fluxes for our possible AGN agree to within $\sim$10\%.  If we simply scale this distribution to a circular area with 20$'$ radius, then we expect 635 AGN with fluxes $>$1.4$\times 10^{-15}$ erg cm$^{-2}$ (equivalent to luminosities $> 10^{35}$ erg s$^{-1}$ for sources in M31); this is a factor $\sim$3 times greater than the number of observed AGN candidates, and more than total of observed X-ray sources. The origin of this discrepancy could be astronomical, if absorbing material in the M31 bulge reduces the observed number of faint AGN. Alternatively, this effect could be instrumental, since the sensitivity of the Chandra images is reduced at large off-axis angles by vignetting and reduced exposure; our normalization of the AGN flux distribution assumes uniform sensitivity.

Therefore, we compared the AGN candidates with the expected AGN within 6$'$ of the M31 mucleus. We expected 56 AGN, and observed 66 AGN candidates within this region.  Given that we see an excess of sources consistent with AGN over the expected amount within 6$'$ of M31*, and only $\sim$30\% expected sources within 20$'$ of M31*, this deficit is probably dominated by instrumental effects.

We compare the  0.3--10 keV flux distribution of the observed AGN candidates with the expected 0.5--10 keV flux distribution found by \citet{georgakakis08} for the two regions (with radius 20$'$ and 6$'$) in Fig.~\ref{rank0lf}. Solid, stepped lines indicate the observed flux distributions, while dashed lines show the expected distributions; black lines correspond to the 20$'$ region, and grey lines to the 6$'$ region. We see that the flux distributions of possible AGN are considerably flatter than expected; together with the excess seen in the population within 6$'$, these results suggest that some of the sources with Rank $<$2.6 are unidentified XBs.

\subsection{Number density of XBs and AGN vs distance from M31*}

\begin{figure}
\epsscale{1}
\plotone{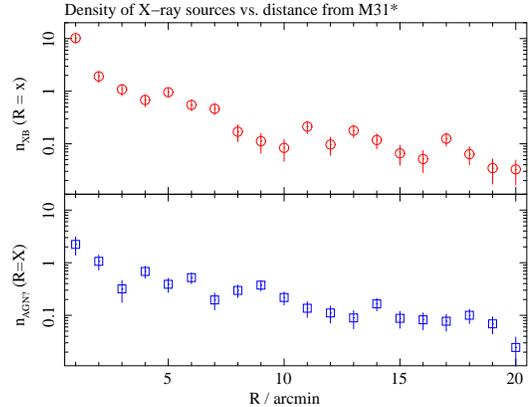}
\caption{Number densities of our XBs (top) and AGN (bottom) vs. distance from galactic center (in the plane of the detector).  We expect XB density to follow stellar mass, while AGN density is expected to be constant. }\label{axvr}
\end{figure}

We calculated the number densities of our 250 XBs  and our 203 possible  AGN per arcmin$^2$  vs. distance from M31* in the detector plane. We present our findings in Fig.~\ref{axvr}.

Our variability survey makes no distinction between low mass X-ray binaries (LMXBs) and high mass X-ray binaries (HMXBs). LMXBs are old, and their numbers tend to follow stellar mass; HMXBs are young, and their populations depend on the star formation rate \citep[see e.g][]{grimm03}. 
A proper comparison between the distribution of our XBs and the expected distribution is beyond the scope of this paper. However, the observed  distribution does appear to agree with expectations \citep[c.f. Fig. 9 of][ taking into account differences in normalization due to different luminosity limits]{voss07}.

We might expect the AGN density to remain constant, with a possible decrease at large distances to account for decreasing instrument sensitivity. However, the number density of possible AGN is considerably higher for regions closer to M31*, again consistent with the presence of some XBs in our sample of possible AGN. 


\subsection{A newly identified  transient black hole candidate in the globular cluster  B128}

In \citet{barnard13}, we identified 26 new black hole candidates (BHCs) in our Chandra observations using a three step process. First we established that the X-ray source was an XB, using SF analysis or by associations with a globular cluster, or by flux. Next we looked for high quality  hard state spectra (represented by power law emission with photon index $\Gamma$ $\sim$ 1.4--2.1) at luminosities above the threshold for neutron star (NS) XBs ($\sim$3$\times 10^{37}$ erg s$^{-1}$).  Finally we modelled these spectra with disk blackbody  + blackbody emission models, and compared the values with the parameter space defined by soft NS XB spectra \citep{lin07,lin09,lin12}.

\begin{figure}
\epsscale{1}
\plotone{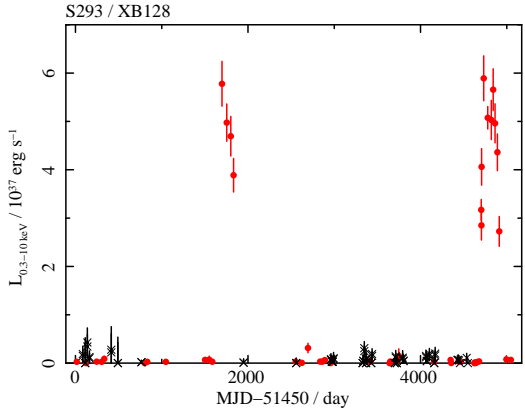}
\caption{Our long-term 0.3--10 keV luminosity lightcurve for S293 / XB128, a recurrent transient X-ray binary associated with the M31 GC B128 that may contain a black hole. ACIS and HRC observations are represented by circles and crosses respectively. }\label{xb128lc}
\end{figure}

S293 in our survey is a newly identified  BHC associated with the confirmed old M31 GC known as B128, following the Revised Bologna Catalog v 3.4 \citep{galleti04,galleti06,galleti07,galleti09}. S293, also known as XB128 following our naming convention for X-ray sources associated with GCs \citep[see e.g.][]{barnard13}, is a transient X-ray source that has exhibited two outbursts during our monitoring campaign. We present a 0.3--10 keV lightcurve of XB128 in Fig.~\ref{xb128lc}.


The first outburst occurred in 2004 May, reaching a 0.3--10 keV luminosity of 5.3$\pm$0.4$\times 10^{37}$ erg s$^{-1}$ and lasted at least 134 days; unfortunately, the nearest observation containing XB128  prior to the burst was 113 days earlier, and the next one was 114 days after the last bright observation. The first outburst was only seen in decline. 

We caught the second outburst during the rise in 2012 August, with a maximum 0.3--10 keV luminosity of 5.9$\pm$0.4$\times 10^{37}$ erg s$^{-1}$; the true peak may not have been observed. It was still active 210 days after our first detection of the outburst, but had disappeared the next time that region was observed 62 days later.

During a $\sim$40 ks ACIS-S observation (Obs ID 14196), a power law fit to the spectrum of XB128 gave $N_{\rm H}$ = 8$\pm$5$\times 10^{20}$ atom cm$^{-2}$ and $\Gamma$ = 1.54$\pm$0.09 with $\chi^2$/dof = 56/63; the 0.3--10 keV luminosity was 5.1$\pm$0.2$\times 10^{37}$ erg s$^{-1}$. This is the best spectrum we have for XB128 that supports a BHC classification, with 1550 net source counts. 

We fitted this spectrum with a disk blackbody  + blackbody emission model, and found the 2--10 keV luminosities for each component. This is for comparison with the NS XB soft state parameter space that was ascertained from hundreds of RXTE spectra analyzed by Lin et al. (2007, 2009, 2012). 

Our criteria for identifying black hole candidates from their disk blackbody  + blackbody fits are presented in \citet{barnard13} and summarized as follows. Lin et al. (2007, 2009) were unable to obtain succesful fits to hard state spectra with their double thermal  (disk blackbody  + blackbody) model; the disk blackbody temperature is forced to low values because it must account for the low energy flux, and contributes little to the 2--10 keV flux. By contrast, Lin et al. (2007, 2009, 2012) found the the disk blackbody component dominated their bright NS spectra, with temperatures $\ga$ 1 keV. Since we identify our BHCs from high luminosity hard states, we expect double thermal fits to BHC spectra to result in low temperatures and small  contributions to the 2--10 keV flux from the disk blackbody components \citep{barnard13}.


\begin{figure}
\epsscale{1}
\plotone{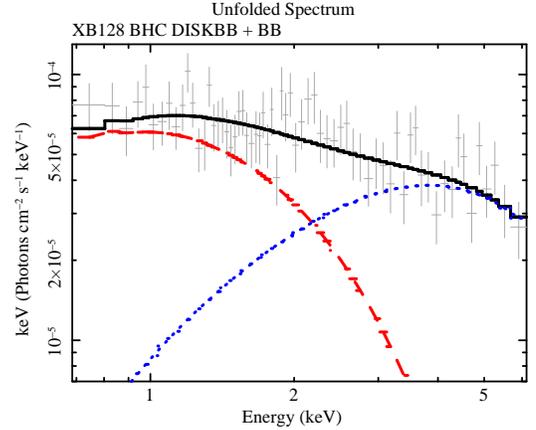}
\caption{Unfolded  spectrum for S293 / XB128 from the $\sim$40 ks ACIS-S observation 14196, with the best fit double thermal model. The disk blackbody and blackbody components are respresented by dashed and dotted curves respectively.  The y-axis is multiplied by channel energy.  }\label{xb128spec}
\end{figure}

When fitting the double thermal model, the absorption was fixed to 7$\times 10^{20}$ atom cm$^{-2}$, equivalent to the Galactic line-of-sight absorption; the absorption was unreasonably low when freely fitted.  The disk blackbody contributed 4$\pm$2$\times 10^{36}$ erg s$^{-1}$ to the 2--10 keV luminosity, with an inner disk temperature of 0.59$_{-0.11}^{ +0.16}$ keV. The blackbody component had a temperature of 1.35$_{-0.18}^{ +0.4}$ keV, emitting 2.4$\pm$0.2$\times 10^{37}$ erg s$^{-1}$. The total 2--10 keV luminosity was 2.8$\pm$0.3$\times 10^{37}$ erg s$^{-1}$, and  $\chi^2$/dof = 57/62. We present the unfolded spectrum fitted with this model in Fig.~\ref{xb128spec}. The disk blackbody temperature was 2.8$\sigma$ below 1 keV; furthemore, its 2--10 keV contribution was 3.8$\sigma$ below 50\%. Hence, XB128 is unlikely to contain a NS primary (probability of disk blackbody temperature and contribution being consistent with a NS $\sim$7$\times 10^{-7}$), and is a strong black hole candidate.

Optical spectroscopy of the globular cluster B128 reavealed that it has an age  $\sim$10$^{10.1}$ year, a mass of $\sim10^{5.32}$ M$_{\odot}$, and metallicity [Fe/H] = $-$0.56; it is ranked 68 out of 379 GCs by metallicity ($>$82\%) and ranked 204 ($>$46\%) by mass  \citep{caldwell09,caldwell11}. We have previously found 11 GC BHCs, and they are all found in GCs that are significantly more massive, more metal rich, or both, than the general GC population in M31 \citep{barnard08,barnard09, barnard11b, barnard2012c, barnard13}.  The GC B128 continues this trend, being near average mass but particularly metal rich.

\section{Summary and conclusions} 
We have applied a novel technique for distinguishing between XBs and AGN to  528 X-ray sources in the central region of M31, observed $\sim$170 times over the last 13 years with Chandra. We identified 250 likely XBs in total; this includes $\sim$200 new X-ray binary identifications. Previously, only $\sim$50 XBs and $\sim$40 XB candidates had been identified out of $\sim$2000 X-ray sources within the $D_{25}$ region of M31 \citep{stiele11}.

We find that up to 205 X-ray sources with 0.3--10 keV fluxes $\ga$1.4$\times 10^{-15}$ erg cm$^{-2}$ s$^{-1}$  are consistent with being AGN; this is a factor $\sim$3 lower than expected from the 0.5--10 keV flux distribution found by \citet{georgakakis08}. We find that this difference is due to instrumental effects: simply scaling the  \citet{georgakakis08} relation by  by area implies a constant sensitivity, but the Chandra sensitivity significantly decreases with increasing off-axis angle. 

Our SF analysisis technique for identifying X-ray binaries could be applied to several well-studied nearby galaxies. We find that the low luminosity XBs tend to be significantly more variable than the high luminosity ones \citep[in agreement with observations of Galactic XB behavior][]{muno02}; hence this technique is not suitable for distant galaxies where only very luminous XBs are detected.

Furthermore, we identify a new transient black hole candidate, associated with the confirmed globular cluster B128. This X-ray source, known as XB128, exhibited two outbursts during our monitoring observations. We obtained a 40 ks ACIS-S spectrum during the observed peak of the second outburst, and fitted it with a disk blackbody + blackbody emission model for comparison with the full gammut of specra exhibited by Galactic neutron star binaries and analyzed by Lin et al. (2007, 2009, 2012). The disk blackbody component for XB128 was inconsistent with neutron star spectra at  a  $>$5$\sigma$ level, hence we label it a black hole candidate, following \citet{barnard13}.

\section*{Acknowledgments}
We thank the anonymous referee, whose thoughtful comments substantially  improved this paper. This research has made use if data obtained from the Chandra data archive, and software provided by the Chandra X-ray Center (CXC). This work was supported by Chandra grants GO2-13106X, and GO3-14095X.




{\it Facilities:} \facility{CXO (ASIS)} \facility{CXO (HRC)}.





\clearpage






\addtocounter{table}{-2}
\begin{table*}
\begin{center}
\caption{continued. }
\renewcommand{\arraystretch}{.9}
\begin{tabular}{cccccccccccc}
\tableline\tableline
Src     &     Pos     &    $\sigma_{\rm RA}$/$''$     &    $\sigma_{\rm Dec}$/$''$     &      $O_{\rm A}$     &     $O_{\rm H}$    &     Properties    &     $L_{\rm con}$ /10$^{37}$     &     $\chi^2$/dof     &     SF Rank\\
\tableline 
61 & 00:41:59.30 +41:05:39.0 &  &  & 15 & 0 & NE=X & 0.09$\pm$0.02 &  7/14 & 0.0$^g$\\
62 & 00:41:59.43 +41:20:46.9$^a$ & 1.5$^a$ & 1.5$^a$ & 39 & 39 & H & 0.072$\pm$0.009$^d$ &  50/77 & 0.0$^g$\\
63 & 00:42:00.58 +41:14:08.7 & 0.6 & 0.8 & 52 & 27 & NE & 0.041$\pm$0.006 &  63/78 & 0.0$^g$\\
64 & 00:42:00.78 +41:16:31.9 & 0.7 & 0.8 & 52 & 33 & NE & 0.038$\pm$0.006 &  45/84 & 0.0$^g$\\
65 & 00:42:02.83 +41:27:57.6$^a$ & 1.8$^a$ & 1.8$^a$ & 6 & 31 & NC & 0.066$\pm$0.014 &  26/36 & 0.6$^g$\\
66 & 00:42:02.93 +41:23:15.5$^a$ & 1.9$^a$ & 1.9$^a$ & 25 & 37 & H & 0.109$\pm$0.012$^d$ &  60/61 & 3.3$^f$\\
67 & 00:42:04.12 +41:09:29.4 & 0.6 & 0.6 & 32 & 53 & NE & 0.131$\pm$0.012$^d$ &  90/84 & 0.1$^g$\\
\end{tabular}

\end{center}
\end{table*}

\addtocounter{table}{-1}
\begin{table*}
\begin{center}
\caption{continued. }
\renewcommand{\arraystretch}{.9}
\begin{tabular}{cccccccccccc}
\tableline\tableline
Src        &        Pos        &       $\sigma_{\rm RA}$/$''$        &       $\sigma_{\rm Dec}$/$''$        &         $O_{\rm A}$        &        $O_{\rm H}$       &        Properties       &        $L_{\rm con}$ /10$^{37}$        &        $\chi^2$/dof        &        SF Rank\\
\tableline
131 & 00:42:20.43 +41:13:13.7 & 0.5 & 0.4 & 69 & 38 & NE=X & 0.041$\pm$0.006 &  66/106 & 0.0$^g$\\
132 & 00:42:20.56 +41:26:39.3 & 0.5 & 0.5 & 14 & 58 & NE=X & 0.366$\pm$0.013 &  76/71 & 6.4$^f$\\
133 & 00:42:20.63 +40:57:13.4 & 0.9 & 0.7 & 4 & 7 & NE & 0.16$\pm$0.05$^c$ &  4/3 & 0.6$^g$\\
134 & 00:42:20.68 +41:16:07.8 & 0.4 & 0.4 & 59 & 28 & NE=T & 0.021$\pm$0.003 &  83/86 & 3.1$^f$\\
135 & 00:42:20.80 +41:13:44.1 & 0.4 & 0.3 & 69 & 34 & NE & 0.036$\pm$0.005 &  56/102 & 0.0$^g$\\
136 & 00:42:20.82 +41:17:57.3 & 0.6 & 0.6 & 39 & 41 & NE & 0.020$\pm$0.003 &  69/79 & 1.7$^g$\\
137 & 00:42:20.90 +41:15:20.1 & 0.5 & 0.4 & 40 & 0 & NE=T & 0.055$\pm$0.010 &  140/39 & 4.1$^f$\\
\end{tabular}

\end{center}
\end{table*}

\addtocounter{table}{-1}
\begin{table*}
\begin{center}
\caption{continued. }
\renewcommand{\arraystretch}{.9}
\begin{tabular}{cccccccccccc}
\tableline\tableline
Src        &        Pos        &       $\sigma_{\rm RA}$/$''$        &       $\sigma_{\rm Dec}$/$''$        &         $O_{\rm A}$        &        $O_{\rm H}$       &        Properties       &        $L_{\rm con}$ /10$^{37}$        &        $\chi^2$/dof        &        SF Rank\\
\tableline 
201 & 00:42:35.232 +41:20:05.86 & 0.11 & 0.10 & 108 & 61 & H & 1.707$\pm$0.019 &  907/168 & 33.8$^f$\\
202 & 00:42:35.88 +41:10:07.9 & 0.6 & 0.6 & 74 & 38 & H & 0.038$\pm$0.005$^d$ &  74/111 & 0.0$^g$\\
203 & 00:42:36.06 +41:20:21.8 & 0.6 & 0.5 & 56 & 41 & NE & 0.025$\pm$0.004 &  72/96 & 0.0$^g$\\
204 & 00:42:36.33 +40:58:48.0 & 0.8 & 0.8 & 11 & 14 & H & 0.28$\pm$0.05 &  28/24 & 4.4$^f$\\
205 & 00:42:36.611 +41:13:50.24 & 0.14 & 0.10 & 98 & 56 & [*]=var* & 0.095$\pm$0.004$^e$ &  581/153 & 7.9\\
206 & 00:42:37.05 +41:14:28.2 & 0.4 & 0.3 & 52 & 41 & NE=[Nova] & 0.021$\pm$0.003 &  1063/92 & 175.4\\
207 & 00:42:37.88 +41:21:33.9 & 0.5 & 0.5 & 72 & 31 & NE & 0.040$\pm$0.004 &  77/102 & 1.3$^g$\\
\end{tabular}

\end{center}
\end{table*}

\addtocounter{table}{-1}
\begin{table*}
\begin{center}
\caption{continued. }
\renewcommand{\arraystretch}{.9}
\begin{tabular}{cccccccccccc}
\tableline\tableline
Src        &        Pos        &       $\sigma_{\rm RA}$/$''$        &       $\sigma_{\rm Dec}$/$''$        &         $O_{\rm A}$        &        $O_{\rm H}$       &        Properties       &        $L_{\rm con}$ /10$^{37}$        &        $\chi^2$/dof        &        SF Rank\\
\tableline 
271 & 00:42:45.230 +41:17:22.38 & 0.11 & 0.10 & 96 & 52 & T & 0.190$\pm$0.005 &  4052/147 & $>$320$^f$\\
272 & 00:42:45.237 +41:16:11.12 & 0.11 & 0.10 & 109 & 62 & NE=X & 0.266$\pm$0.005 &  679/170 & 27.6$^f$\\
273 & 00:42:45.526 +41:16:52.60 & 0.4 & 0.3 & 66 & 39 & NE & 0.026$\pm$0.004 &  53/104 & 0.5$^g$\\
274 & 00:42:45.598 +41:16:08.59 & 0.11 & 0.10 & 109 & 61 & NE=X & 0.518$\pm$0.009 &  1594/169 & $>$320$^f$\\
275 & 00:42:45.84 +41:24:33.4 & 0.5 & 0.5 & 74 & 35 & H & 0.111$\pm$0.008 &  67/108 & 1.2$^g$\\
276 & 00:42:45.946 +41:10:36.53 & 0.12 & 0.10 & 56 & 43 & NE=BHC,T & 0.069$\pm$0.006 &  8381/98 & 118.4$^f$\\
277 & 00:42:46.007 +41:16:19.59 & 0.11 & 0.10 & 97 & 49 & NE=X,T & 0.138$\pm$0.004 &  1981/145 & 4.3$^f$\\
\end{tabular}

\end{center}
\end{table*}

\addtocounter{table}{-1}
\begin{table*}
\begin{center}
\caption{continued. }
\renewcommand{\arraystretch}{.9}
\begin{tabular}{cccccccccccc}
\tableline\tableline
Src         &         Pos         &        $\sigma_{\rm RA}$/$''$         &        $\sigma_{\rm Dec}$/$''$         &          $O_{\rm A}$         &         $O_{\rm H}$        &         Properties        &         $L_{\rm con}$ /10$^{37}$         &         $\chi^2$/dof         &         SF Rank\\
\tableline 

341 & 00:42:53.68 +41:25:50.9 & 0.5 & 0.5 & 46 & 62 & !SNR & 0.337$\pm$0.009 &  442/107 & 3.3$^f$\\
342 & 00:42:53.7 +41:10:59$^a$ & 7$^a$ & 7$^a$ & 61 & 31 & [SSS] & 0.035$\pm$0.005 &  56/91 & 0.0\\
343 & 00:42:54.35 +41:13:33.9 & 0.4 & 0.4 & 54 & 33 & NE & 0.017$\pm$0.003 &  63/86 & 0.0$^g$\\
344 & 00:42:54.54 +41:04:00.5 & 0.6 & 0.7 & 11 & 38 & NE=X & 0.067$\pm$0.015 &  36/48 & 3.9$^f$\\
345 & 00:42:54.935 +41:16:03.19 & 0.11 & 0.10 & 109 & 62 & H=BHC & 3.12$\pm$0.03$^c$ &  5782/108 & $>$320$^f$\\
346 & 00:42:55.18 +41:20:45.1$^a$ & 1.4$^a$ & 1.4$^a$ & 33 & 50 & [SSS] & 0.020$\pm$0.003$^d$ $^e$ &  173/82 & 126.3\\
347 & 00:42:55.197 +41:18:36.12 & 0.12 & 0.10 & 109 & 62 & NE & 0.417$\pm$0.007 &  605/170 & 133.1$^f$\\
\end{tabular}

\end{center}
\end{table*}

\addtocounter{table}{-1}
\begin{table*}
\begin{center}
\caption{continued. }
\renewcommand{\arraystretch}{.9}
\begin{tabular}{cccccccccccc}
\tableline\tableline
Src        &        Pos        &       $\sigma_{\rm RA}$/$''$        &       $\sigma_{\rm Dec}$/$''$        &         $O_{\rm A}$        &        $O_{\rm H}$       &        Properties       &        $L_{\rm con}$ /10$^{37}$        &        $\chi^2$/dof        &        SF Rank\\
\tableline 
411 & 00:43:09.866 +41:19:00.76 & 0.14 & 0.10 & 91 & 55 & T=[GC],BHC & 0.120$\pm$0.006 &  2667/145 & $>$320$^f$\\
412 & 00:43:09.97 +41:23:32.8 & 0.3 & 0.4 & 46 & 39 & XB,T & 0.042$\pm$0.005 &  282/84 & 23.5$^f$\\
413 & 00:43:10.07 +41:27:19.0 & 0.9 & 0.7 & 18 & 33 & H & 0.083$\pm$0.012 &  32/50 & 4.4$^f$\\
414 & 00:43:10.11 +41:17:20.3 & 0.5 & 0.7 & 60 & 29 & [GC] & 0.024$\pm$0.005 &  66/88 & 1.6$^f$\\
415 & 00:43:10.622 +41:14:51.30 & 0.11 & 0.10 & 103 & 62 & H=GC,BHC & 12.936$\pm$0.106$^c$ &  511/102 & 25.8$^f$\\
416 & 00:43:10.78 +41:30:43.0$^a$ & 1.1$^a$ & 1.1$^a$ & 10 & 36 & H & 0.064$\pm$0.015 &  38/45 & 0.0$^g$\\
417 & 00:43:11.31 +41:17:14.0 & 0.6 & 0.6 & 66 & 35 & NE & 0.030$\pm$0.005 &  70/100 & 0.0$^g$\\
\end{tabular}

\end{center}
\end{table*}

\addtocounter{table}{-1}
\begin{table*}
\begin{center}
\caption{continued. }
\renewcommand{\arraystretch}{.9}
\begin{tabular}{cccccccccccc}
\tableline\tableline
Src        &        Pos        &       $\sigma_{\rm RA}$/$''$        &       $\sigma_{\rm Dec}$/$''$        &         $O_{\rm A}$        &        $O_{\rm H}$       &        Properties       &        $L_{\rm con}$ /10$^{37}$        &        $\chi^2$/dof        &        SF Rank\\
\tableline 
481 & 00:43:33.72 +41:14:06.30 & 0.7 & 0.9 & 38 & 34 & H & 0.100$\pm$0.012 &  60/71 & 0.0$^g$\\
482 & 00:43:33.86 +41:08:21.40 & 1.3 & 1.0 & 11 & 27 & NE & 0.112$\pm$0.015$^d$ &  42/37 & 5.2$^f$\\
483 & 00:43:34.29 +41:02:07.80 & 0.6 & 0.6 & 8 & 8 & H & 0.12$\pm$0.04$^d$ &  11/15 & 5.0$^f$\\
484 & 00:43:34.33 +41:13:23.10 & 0.2 & 0.2 & 50 & 62 & H=BHC & 3.77$\pm$0.04 &  1450/111 & 35.7$^f$\\
485 & 00:43:36.52 +41:09:25.20 & 0.9 & 1.4 & 14 & 34 & NE & 0.089$\pm$0.015$^d$ &  39/47 & 2.7$^f$\\
486 & 00:43:36.77 +41:08:12.00 & 1.3 & 1.0 & 9 & 38 & GC & 0.016$\pm$0.004 &  52/46 & 5.4$^f$\\
487 & 00:43:37.32 +41:14:43.10 & 0.2 & 0.2 & 40 & 62 & GC=BHC & 6.19$\pm$0.09$^c$ &  100/39 & 0.0$^f$\\
\end{tabular}

\end{center}
\end{table*}


\begin{table*}
\begin{center}
\caption{continued. }
\renewcommand{\arraystretch}{.9}
\begin{tabular}{cccccccccccc}
\tableline\tableline
Src    &    Model    &    $N_{\rm H}$/10$^{22}$    &    $\chi^2$/dof    &    Param     &    $\chi^2$/dof \\
\tableline 
265 & PO & 0.137 $\pm$ 0.015 & 10/16 & 2.08 $\pm$ 0.04 & 11/17 \\
266 & PO & 0.09 $\pm$ 0.03 & 1.4/7 & 1.60 $\pm$ 0.09 & 5/7 \\
267 & PO & 0.06 $\pm$ 0.06 & --- & 1.78 $\pm$ 0.19 & --- \\
268 & PO & 0.1 $\pm$ 0.10 & --- & 2.7 $\pm$ 0.6 & --- \\
269 & PO & 0.106 $\pm$ 0.017 & 8/11 & 1.78 $\pm$ 0.05 & 10/11 \\
270 & PO & 0.3 $\pm$ 0.2 & 0.8/2 & 1.7 $\pm$ 0.4 & 0.4/2 \\
271 & PO & 0.23 $\pm$ 0.10 & --- & 2 $\pm$ 0.2 & --- \\
\end{tabular}

\end{center}
\end{table*}


\end{document}